\documentclass[aps,prc,twocolumn,superscriptaddress,amsmath,showpacs,amssymb,floatfix]{revtex4-1}
\usepackage{graphicx}
\usepackage{color}
\usepackage{ulem}
\usepackage{amsmath}

\newcommand\bp{{\bf p}}

\newcommand\bq{{\bf q}}
\newcommand\br{{\bf r}}
\newcommand\rB{{\rm B}}
\newcommand\rI{{\rm I}}
\newcommand\rBB{{\rm BB}}
\newcommand\rIB{{\rm IB}}
\newcommand\rex{{\rm ex}}

\newcommand{\lk}{\left(}
\newcommand{\rk}{\right)}
\newcommand{\ltk}{\left\{}
\newcommand{\rtk}{\right\}}
\newcommand{\ldk}{\left[}

\newcommand{\rdk}{\right]}
\newcommand\beq{ \begin{eqnarray} }
\newcommand\eeq{ \end{eqnarray} }

\newcommand{\intr}{\int_{\bf r}}

\newcommand{\intdbr}{\int\! d^3\br\;}

\begin{document}

\title{Bose polaron in spherical trap potentials: \\
Spatial structure and quantum depletion}
\author{J.~Takahashi}
\email{takahashi.j@aoni.waseda.jp}
\affiliation{Department of Electronic and Physical Systems, Waseda University,
Tokyo 169-8555, Japan}
\author{R.~Imai}
\email{rimai@asagi.waseda.jp}
\affiliation{Department of Materials Science, Waseda University,
Tokyo 169-8555, Japan}
\author{E.~Nakano}
\email{e.nakano@kochi-u.ac.jp}
\affiliation{Department of Mathematics and Physics, Kochi University,
Kochi 780-8520, Japan}
\author{K.~Iida}
\email{iida@kochi-u.ac.jp}
\affiliation{Department of Mathematics and Physics, Kochi University,
Kochi 780-8520, Japan}

\date{\today}

\begin{abstract}
We investigate how the presence of a localized impurity in a Bose-Einstein
condensate of trapped cold atoms that interact with each other weakly and
repulsively affects the profile of the condensed and excited components at
zero temperature.  By solving the Gross-Pitaevskii and Bogoliubov-de Gennes
equations, we find that an impurity-boson contact attraction (repulsion)
causes both components to change in spatial structure in such a way as to be
enhanced (suppressed) around the impurity, while slightly declining (growing)
in a far region from the impurity.  Such behavior of the quantum depletion
of the condensate can be understood by decomposing the impurity-induced change
in the profile of the excited component with respect to the radial and
azimuthal quantum number.  A significant role of the centrifugal potential
and the ``hole'' excitation level is thus clarified.
\end{abstract}

\maketitle

\section{Introduction}
Polarons are quasiparticles conceptually well established in solid state
physics.  For instance, a conducting electron in an ionic crystal moves
together with the locally induced polarization to form a polaron that
has the energy spectrum modified from that of an electron in
vacuum \cite{Landau1,FPZ1,LLP1}.  Such a medium-modified
electron serves as an essential building block for more complex many-body phenomena,
e.g., high $T_c$ superconductivity \cite{Mahan1,polaronreview1,polaronreview2}.
Recently cold atomic gas experiments have offered various types of atomic
polarons, i.e., impurities (minority atoms) that are either
immersed in a trapped Bose gas with Bose-Einstein condensation (BEC) or in a
degenerate Fermi gas and eventually dressed by a virtual cloud of
the corresponding particle-hole excitations.  The former and latter
are called Bose polarons
\cite{Catani1,Scelle1,Hohmann1,Compagno1,Jrgensen1,Hu1,Rentrop1} and
Fermi polarons \cite{Schirotzek1,Froehlich1,Kohstall1,Scazza1,Cetina1},
respectively.  At sufficiently low temperatures, the interaction
between impurity and medium atoms is characterized by a low-energy $s$-wave
scattering length, which is in turn tunable by external magnetic fields
as predicted by the Fano-Feshbach theory \cite{PethickSmith1}.  The
polaron energy and spectral strength have been measured by, e.g.,
radio frequency spectroscopy \cite{Jrgensen1,Hu1,Schirotzek1,Froehlich1},
which utilizes the radio absorption probability of impurity atoms
in two different hyperfine states that interact with medium atoms
only weakly and relatively strongly.

The above experiments have also energized theoretical investigations,
which include various issues of atomic polarons in zero-temperature media:
quasiparticle properties of Bose polarons
\cite{Tempere1,Li1,Shashi1,Dehkharghani5,Christensen1,
Vlietinck1,Grusdt1,Grusdt3,Shchadilova1, Blinova1, Drescher1}
and Fermi polarons
\cite{Chevy1,Chevy2,Massignan1,Schmidt1,Koschorreck1,Schmidt6,
Vlietinck2,Trefzger1,Trefzger2,Massignan5,Lan2,Yi1,Kamikado1,Kain1},
the self-localization of impurities in quasi-1D BEC media
\cite{Cucchietti1,Sacha1,Kalas1,Bruderer20,Boudjemaa1},
polaronic spectral changes from a weak to strong coupling regime for
attractive interactions
\cite{Casteels1,Rath1,Ardila2,Schmidt5,Ardila6},
few-body physics around the unitarity limit
\cite{Levinsen2,Levinsen4,Sun1,Naidon1,Shchadilova2},
dynamics of the polaron formation
\cite{Shchadilova2,Parish5,Grusdt6,Ashida1},
and open quantum dynamics
\cite{Lampo5,Nielsen3,Mistakidis9}.
Beside these 'conventional' atomic polarons, more exotic ones have been
proposed, e.g., $p$-wave \cite{Levinsen5} and dipolar-type
\cite{Kain3,Ardila4} polarons,
angulons \cite{Lemeshko3,Lemeshko4,Lemeshko7,Lemeshko8}, and
Bose polarons near the transition temperature \cite{Levinsen8,Guenther8}.
Moreover, thermal evolution of Fermi polarons has been recently in the
recent experiment\cite{Yan8} prior to theoretical investigations
\cite{Tajima1,Tajima2}
and critical properties of
Bose polarons have been experimentlly observed \cite{ZZYan1}.
Whereas most of the theoretical studies
mentioned above assume that the system is spatially uniform, Bose
polarons in rotationally symmetric trap potentials
\cite{Nakano1,Nakano2} that are used in experiments have been
theoretically studied to figure out the ground-state properties
at a given total angular momentum, a conserved quantity in the
system considered here.
Note that the total momentum is the corresponding
quantity in translationally-symmetric uniform systems.

Up to now, however, theoretical studies of Bose polarons
have yet to reveal the detailed structures of the condensate
and fluctuations in the presence of impurity, especially in 3D trapped systems.
This is because the Bogoliubov approximation
has been used basically
within a theoretical framework that assumes that the presence of the
impurity does not change the spatial profile of the condensate nor the
number of condensed bosons.
The Bogoliubov approximation is supposed to be
valid if the fraction of non-condensed bosons that are caused to occur
by all interactions assumed in the system is negligibly small ($\ll 1$).
There are some estimates of this condition with respect to the
boson-boson and the impurity-boson interaction strengths
\cite{Bruderer1,Nakano3}.  In general, a repulsive interaction among
bosons disturbs them in their condensing into a coherent
state by causing fluctuations to generate a finite fraction
of non-condensed bosons even at zero temperature, a phenomenon referred
to as {\it the quantum depletion of the condensate}
\cite{PethickSmith1,Dalfovo1,Pitaevskii1}.
In superfluid $^4$He, for instance, the fraction is no less than
about $80-90$ \% \cite{Legget1}, while in cold-atomic condensates
it is less than $10$ \% because a very weak repulsive interaction can be
naturally realized \cite{Dalfovo2}; it has been recently observed
in the experiment \cite{Lopes1} in a box trap.
Aside from the validity argument of the approximations used
in theoretical studies of Bose polarons, it is interesting to find
a missing piece, namely, to
examine the open question of how the impurity gives a local
feedback to the medium bosons:
The impurity inevitably induces a local deformation of the condensate
and local quantum fluctuations.
Our motivation is to examine such local modifications caused
by an impurity localized in a trap potential.
In the present paper, therefore, we consider a single
Bose polaron in spherically symmetric trap potentials
and figure out details of spatial structure of the condensate
and Bogoliubov excitations by allowing for the changes induced
around the impurity by the impurity-boson interaction.

This paper is organized as follows:  In Sec.~II we set up our system,
write down the effective Hamiltonian in detail, and derive a set of coupled
equations under some reasonable approximations.  In Sec.~III we show
numerical results obtained by solving the equations derived
in Sec.~II.  In Sec.~IV we end up with summary.

\section{Formulation}
In this section, we shall write down
a set of equations that describe
spatial structure of an impurity and
the condensed and excited components of a weakly interacting
Bose gas in trap potentials.

\subsection{Effective Hamiltonian}
We consider the zero-temperature system of a single atomic impurity
immersed in a dilute atomic Bose gas.
The impurity and the gas are trapped
in the confinement potentials $V_\rI(\br)$ and $V_\rB(\br)$.
Bosons are assumed to interact with each other weakly and repulsively,
while the impurity-boson interaction is assumed to be tunable between
positive and negative values using the Fano-Feshbach resonance.
Such a system  can be described by the low energy effective
Hamiltonian, $\mathcal{H}=H_{\rm imp}+H_{\rm B}+H_{\rm int}$, that
is composed of the part of the trapped single impurity,
the part of the trapped boson gas, and the part of the
impurity-boson interaction, respectively,  i.e.,
\beq
H_{\rm imp}
&=&
\frac{\hat{\bp}^2}{2m_\rI}+ \frac{m_\rI \omega_\rI^2}{2} \hat{\bq}^2,
\\
H_\rB
&=&
\intr \hat{\phi}^\dagger(\br)
\ldk
h_\rB
\!+\!\frac{g_\rBB}{2}\hat{\phi}^\dagger(\br)\hat{\phi}(\br)
\!-\!\mu
\rdk
\hat{\phi}(\br),
\\
H_{\rm int}&=&
g_\rIB \intr
\hat{\phi}^\dagger(\br) \delta^{(3)}(\hat{\bq}-\br)\hat{\phi}(\br).
\eeq
Here,
$
h_\rB
=-\frac{\hbar^2 {\nabla}^2}{2 m_\rB}
 + V_\rB(\br)
$, and  we have used the first (second) quantized form for the impurity
(bosons). We have introduced the abbreviated notation for the spacial integral:
$
\intr = \intdbr.
$
For the boson-boson and boson-impurity interaction,
the effective coupling constants and $s$-wave scattering lengths are
related by
\beq
    g_\rBB &=& \frac{4\pi\hbar^2}{m_\rB} a_\rBB, \\
    g_\rIB &=& \frac{2\pi\hbar^2}{m_{\rm red}} a_\rIB,
\eeq
where $m_{\rm red}=m_\rB m_\rI/(m_\rB + m_\rI)$.

We aim to figure out the ground state properties of this system
at zero temperature.  In formulation we first take the expectation of
$\mathcal{H}$ with respect to a normalized impurity state $|{\rm imp}\rangle$
that is yet to be determined. The resultant expression reads
\beq
\mathcal{H}_\rB
&=& \langle \mathcal{H} \rangle_{\rm imp} \notag \\
&=& \langle H_{\rm imp}\rangle_{\rm imp} \!+\! H_\rB
\!+\! g_\rIB \! \intr |\psi(\br)|^2 \hat{\phi}^\dagger(\br) \hat{\phi}(\br),
\label{Hb}
\eeq
where $\psi(\br)=\langle \br|{\rm imp}\rangle$ is the wave function of
the impurity, and
$
 \langle H_{\rm imp}\rangle_{\rm imp}
  = \intr \psi^*(\br)
    \ldk -\frac{\hbar^2 {\nabla}^2}{2 m_\rI}
         +V_\rI(\br)
    \rdk \psi(\br)
$.
The $\mathcal{H}_\rB$ defined above represents an effective Hamiltonian
for bosons in the presence of an additional potential due to the impurity
wave function.

Now we expand the boson field operator into the condensate and
its fluctuations as
$\hat{\phi}(\br) =\phi(\br) + \hat{\varphi}(\br)$,
where $\phi(\br) :=\langle \hat{\phi}(\br) \rangle$,
with $\langle\cdots\rangle$ defined as the expectation value by
{\it vacuum of bosons in the presence of an
impurity} \cite{note},
corresponding to the ground state of the unperturbed Hamiltonian
to be specified below.
This implies $\langle \hat{\varphi}(\br)\rangle=0$.
Expanding $\mathcal{H}_\rB$ with respect to the fluctuations,
we obtain
\beq
{\mathcal H}^{(0)} \!&=&\!
\langle H_{\rm imp}\rangle_{\rm imp} \notag\\
&+&
\intr \! \phi^*
\ldk
h_\rB
+g_\rIB |\psi|^2
+\frac{g_\rBB}{2} |\phi|^2
-\mu
\rdk
\phi,
\\
{\mathcal H}^{(1)} \!&=&\!
\intr \! \hat{\varphi}^\dagger
\lk
h_\rB
\!+\!g_\rIB |\psi|^2
\!+\!g_\rBB |\phi|^2
\!-\!\mu
\rk
\phi
\!+\!{\it h.c.},
\\
{\mathcal H}^{(2)} \!&=&\!
\frac12
\intr \!
\lk
\begin{array}{cc}
\hat{\varphi}^\dag & \hat{\varphi}
\end{array}
\rk
\lk
\begin{array}{cc}
\mathcal{L} & \mathcal{M} \\
\mathcal{M}^* & \mathcal{L}^*
\end{array}
\rk
\lk
\begin{array}{c}
\hat{\varphi} \\ \hat{\varphi}^\dag
\end{array}
\rk,
\\
{\mathcal H}^{(3)} \!&=&\!
g_\rBB
\intr \! \lk \phi  \, \hat{\varphi}^\dagger\hat{\varphi}^\dagger\hat{\varphi}
+
\phi^* \, \hat{\varphi}^\dagger\hat{\varphi}\hat{\varphi}\rk,
\\
{\mathcal H}^{(4)} \!&=&\!
\frac{g_\rBB}{2}
\intr \! \hat{\varphi}^\dagger\hat{\varphi}^\dagger\hat{\varphi}\hat{\varphi}.
\eeq
where
$
 \mathcal{L} = h_\rB + g_\rIB |\psi|^2+2g_\rBB |\phi|^2 -\mu
$
and
$
 \mathcal{M} = g_\rBB \phi^2
$.

Let us now choose ${\mathcal H}^{(1)}+{\mathcal H}^{(2)}$
as the unperturbed Hamiltonian.
Then, we find
${\mathcal H}^{(1)}=0$ from
$i \hbar \partial_t \langle \hat{\varphi} \rangle
= \left\langle \ldk \hat{\varphi}, {\mathcal H}^{(1)}+{\mathcal H}^{(2)}\rdk
  \right\rangle = 0$
in the interaction picture, which leads to
the Gross-Pitaevskii equation \cite{GP}
\beq
\lk
h_\rB + g_\rIB |\psi|^2 + g_\rBB |\phi|^2 - \mu
\rk
\phi=0
\label{op1}.
\eeq
This does not include fluctuation effects explicitly,
but only through the impurity's wave function (\ref{psi1}).
In fact, Eq.\ (\ref{op1}) is equivalent to the stationary condition from
${\mathcal H}^{(0)}$.
When higher order fluctuation effects
from ${\mathcal H}^{(3)}$ and ${\mathcal H}^{(4)}$ are negligible
due to sufficiently weak interactions, i.e., ${\mathcal H}^{(3)}$ and
${\mathcal H}^{(4)}$ terms in the Hamiltonian can be neglected, then
the Hamiltonian approximately reads
\beq
{\mathcal H}_\rB \simeq {\mathcal H}^{(0)}+  {\mathcal H}^{(2)}.
\eeq
The ground and excited states for the above approximate Hamiltonian
can be determined solely from the diagonalization of ${\mathcal H}^{(2)}$
once the condensation profile is obtained from (\ref{op1}).
In the Bogoliubov representation
$
\hat{\varphi}(\br)=\sum_i [u_i(\br) \hat{\alpha}_i+v_i^*(\br) \hat{\alpha}_i^\dagger],
$
where the operators $\hat{\alpha}_i$ satisfy
the canonical commutation relation:
$[\hat{\alpha}_i, \hat{\alpha}^\dagger_j]=\delta_{ij}$,
we thus obtain the Bogoliubov-de Gennes (BdG) equations
\cite{Bogoliubov, deGennes},
\beq
\lk
\begin{array}{cc}
\mathcal{L} & \mathcal{M} \\
-\mathcal{M}^* & -\mathcal{L}
\end{array}
\rk
\lk
\begin{array}{c}
u_i\\
v_i
\end{array}
\rk
= E_i
\lk
\begin{array}{c}
u_i\\
v_i
\end{array}
\rk,
\label{exc1}
\eeq
where $E_i$ is the eigen-energy of $\mathcal{H}^{(2)}$,
 i.e., the boson excitation energy.  We remark that these
excited bosons can be interpreted as phonons, whose bilinear
coupling with the impurity is similar to the Fr\"{o}hlich-type
electron-phonon coupling in polar semiconductors.

The variational condition with respect to the impurity state
$
 \delta\langle {\mathcal H}_\rB\rangle/\delta \psi^*=0
$
leads to another equation:
\beq
\lk
 h_\rI + g_\rIB|\phi|^2
 + g_\rIB \langle\hat{\varphi}^\dagger\hat{\varphi} \rangle
\rk \psi =0,
\label{psi1}
\eeq
where the expectation is taken
with respect to the bosonic ground state of ${\mathcal H}^{(2)}$,
i.e., the Fock vacuum of $\hat{\alpha}_i$.

Putting Eqs.~(\ref{op1}), (\ref{exc1}), and (\ref{psi1}) together,
we obtain a set of the equations to be solved simultaneously.
We note that the chemical potential $\mu$ will be determined from
\beq
N_\rB= \intr \langle \hat{\phi}(\br)^\dagger \hat{\phi}(\br) \rangle
    := N_0 + N_{\rex},
\eeq
where $N_0$ is the number of the bosons in the condensate,
while $N_{\rex}$  is the number of the bosons in the excited states.
At zero temperature, $N_{\rex}$ represents a depletion of the
condensate \cite{Dalfovo2}, because all the bosons are in the
condensate for a non-interacting, impurity-free system.
The depletion of the condensate can be described by
\beq \label{QD1}
N_{\rex} = \sum_i \intr |v_i(\br)|^2,
\eeq
under the condition that the total number of bosons $N_B$
is kept constant by $\mu$.

\subsection{Spherical trap potentials and further approximations}
Hereafter, for theoretical simplicity, we assume that
the impurity and the gas are trapped in the respective spherical
harmonic-oscillator potentials whose centers coincide:
$V_{\rB}(\br)=\frac{m_\rB \omega_\rB^2}{2}{\br}^2$ and
$V_{\rI}(\br)=\frac{m_\rI \omega_\rI^2}{2}{\br}^2$.
This theoretical setup is motivated by experiments \cite{Hu1},
in which the optical and magneto-optical traps are well described by the
harmonic-oscillator potentials.

In this work, we focus on the quantum depletion of the condensate
in the presence of an impurity.
For simplicity, we ignore the dynamical kickback to the impurity from
the medium Bose gas and consider a situation where the impurity is
strongly bounded by the spherical harmonic trap potential,
i.e., $\hbar\omega_\ell \gg |g_{\rIB}\bar{n}_{\rB}|$,
with $\bar{n}_\rB$ being an averaged boson density
$
\bar{n}_{\rB}
 = N_{\rB} \Big/ \lk \frac{4\pi}{3} d_\rB^3 \rk
$
where $d_{\rm B}=\sqrt{\hbar/m_{\rB}\omega_{\rB}}$.
In this situation, because of the wide energy gap between
the ground and other excited states, the wave function
of the impurity remains in the ground state.
Moreover, the impurity wave function is shrunk by the potential
and feels the kickback from the condensate as the constant potential
which affects only the energy shift; hence, the kickback term from
the condensate in Eq.(\ref{psi1}) can be ignored.
Therefore, we have
\beq
 \psi(\br)
  \simeq \lk \frac{\pi\hbar}{m_\rI \omega_\rI} \rk^{-\frac{3}{4}}
         \exp \lk -\frac{m_\rI \omega_\rI}{2\hbar} r^2 \rk.
\eeq
We note that this condition is inequivalent to the heavy mass limit of
the impurity $m_\rI\gg m_\rB$ because the wave function of the impurity is shrunk
under this condition but the excitation gap remains unchanged.
Since the wave function of the impurity is isotropic, the ground
state of the condensate also has an isotropic form,
\beq
    \phi(\br) = \sqrt{\frac{N_0}{4\pi}} \Phi(r),
\eeq
and the BdG eigenfunctions can be assumed to have a separable form,
\beq
    \ltk
    \begin{matrix}
        u_{n_r \ell m}(\br)\\
        v_{n_r \ell m}(\br)
    \end{matrix}
    \rtk=\ltk
    \begin{matrix}
        \mathcal{U}_{n_r\ell}(r)\\
        \mathcal{V}_{n_r\ell}(r)
    \end{matrix} \rtk
    Y_{\ell m} (\theta_1, \theta_2)  \,,
\eeq
where $(n_r, \, \ell, \, m)$ denote the
radial, azimuthal, and magnetic quantum number, respectively.  Finally,
the depletion of the condensate is characterized by
\beq
    N_\rex &=& \sum_{n_r, \ell} \intr  n_{\rex, n_r\ell}(r),
    \label{QD2}\\
    n_{\rex,n_r\ell}(r) &=& \frac{2\ell+1}{4\pi} |\mathcal{V}_{n_r\ell}(r)|^2.
\eeq

\section{Numerical results and discussion}
Let us now present numerical results for the quantum depletion.
To this end,  we consider a situation in which a $^{40}$K
Fermi impurity is immersed in an $^{87}$Rb Bose condensate,
i.e., $m_\rI/m_\rB \simeq 0.460$.  We set the total number of
the bosons as $N_\rB = 10^5$ and the ratio of the strength of the trap
potentials as $\omega_\rI/\omega_\rB = 5$ with
$\omega_\rB = 20 \times 2\pi \, {\rm Hz}$.
For the boson-boson and
boson-impurity interactions,
we take
$
 1/(a_\rBB \bar{n}_\rB^{1/3}) = 100,
$
and
$
 1/(a_\rIB \bar{n}_\rB^{1/3}) = \pm 10,
$
with
\beq
 \bar{n}_{\rB}
  &=& N_{\rB} \Big/ \lk \frac{4\pi}{3} d_\rB^3 \rk
   = 1.70 \times 10^{15} \,\, {\rm cm}^{-3}, \label{numb} \\
 d_\rB &=& \sqrt{\frac{\hbar}{m_{\rB}\omega_{\rB}}} = 2.41 \times 10^{-4} {\rm cm}.
\eeq
Throughout the numerical calculations
we keep the number of total bosons $N_\rB$ fixed
by tuning the chemical potential.

\subsection{Condensate and depletion in real space}
\begin{figure}[ht]
\begin{center}
\includegraphics[width=7cm]{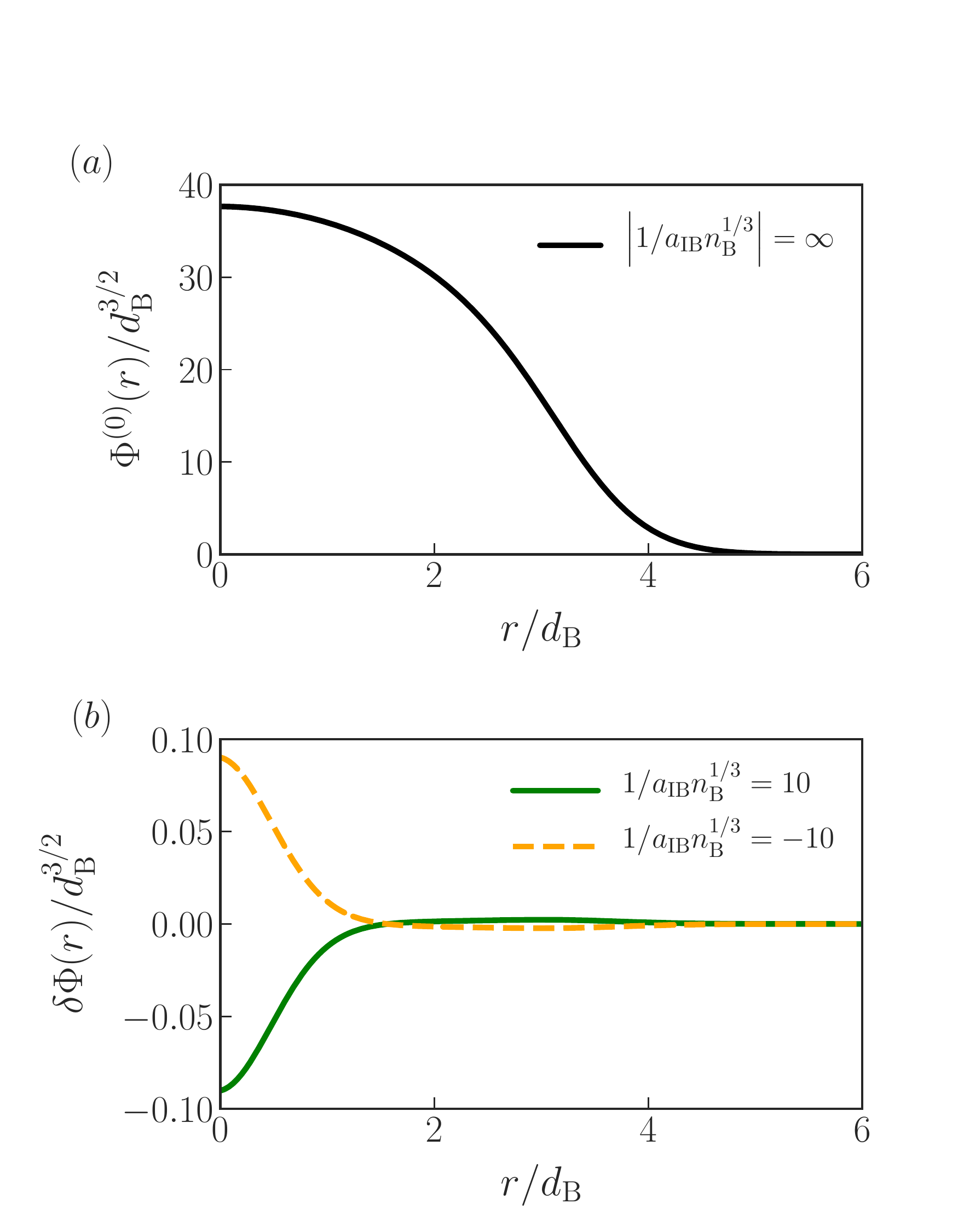}
\caption{
    (Color online)
    (a) Radial  profile of  the order parameter $\Phi(r)$
     in the absence of an impurity and
    (b) difference of  the radial profile of the order
        parameter in the presence of an impurity from the result
        depicted in (a).
}
\label{fig:order_parameter}
\end{center}
\end{figure}

We first consider the spatial dependence of  the condensate
and depletion.  Figure~\ref{fig:order_parameter} shows the radial profile of
the order parameter in the absence of the impurity
$\Phi^{(0)}(r)$ and the change in the order parameter
due to the presence of the impurity, defined by
\beq
    \delta \Phi(r) := \Phi(r) - \Phi^{(0)}(r).
\eeq
Hereafter, we use the superscript $(0)$ that denotes
the absence of the impurity.
Note that the impurity is localized at the center of the trap potential
\cite{com2}.
When the boson-impurity interaction is repulsive $g_{\rIB}>0$
(attractive $g_{\rIB}<0$), therefore,
the condensate feels as if there appeared an additional small
bump (dip) at the center, and eventually is repelled a little bit
from (pulled toward) the impurity as shown in Fig.~\ref{fig:order_parameter}.

\begin{figure}[ht]
\begin{center}
\includegraphics[width=7cm]{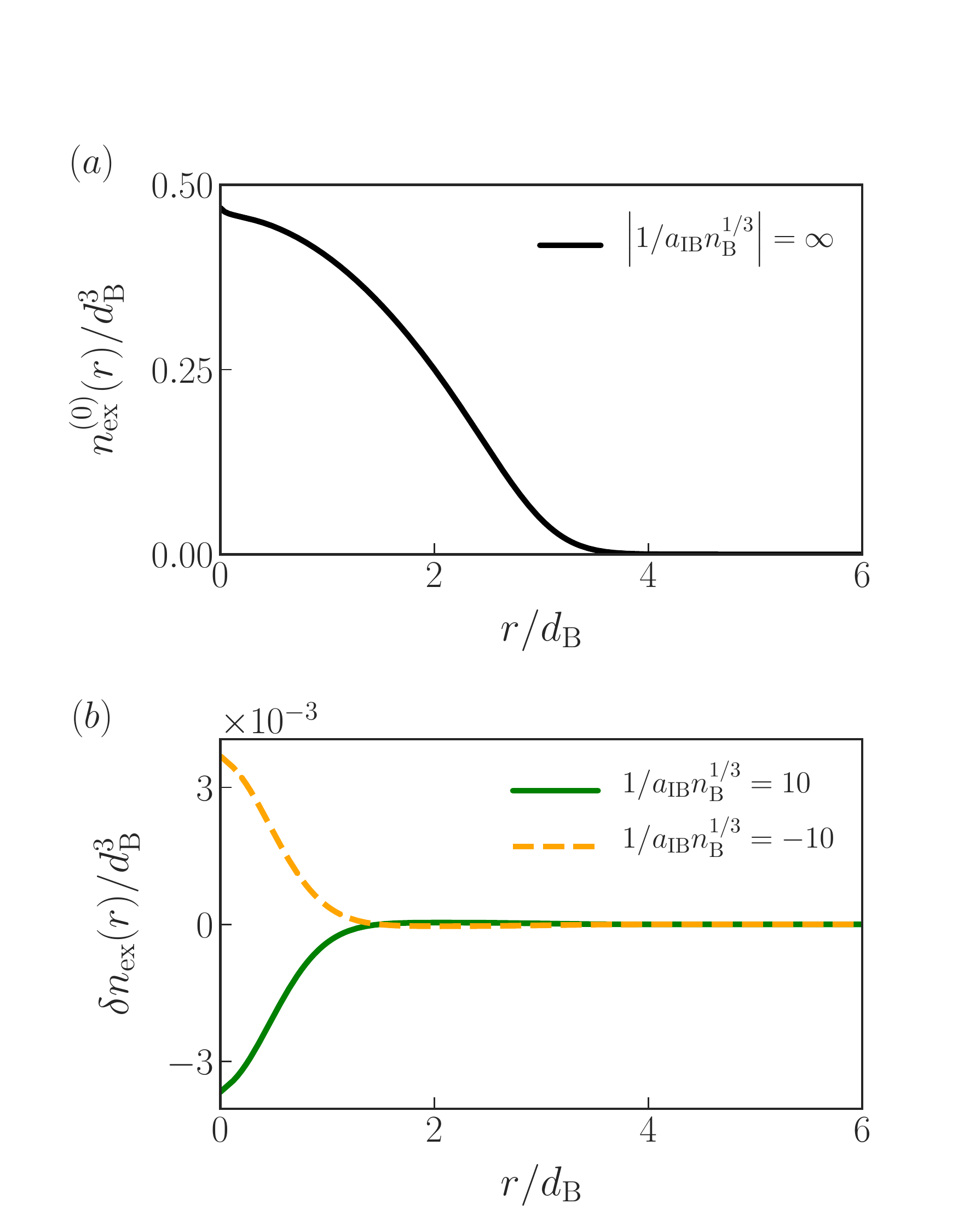}
\caption{
 (Color online)
 Same as Fig.\ \ref{fig:order_parameter} for the density of depletion,
 $\sum_{n_r, \ell} n_{\rex, n_r\ell}(r)$.
}
\label{fig:nex_r}
\end{center}
\end{figure}

Correspondingly, Fig.~\ref{fig:nex_r} exhibits the radial
profile of the density of the quantum depletion,
$\sum_{n_r, \ell} n_{\rex, n_r\ell}(r) := n_\rex (r)$, in the absence of
the impurity and the change in the density of the quantum depletion
due to the presence of the impurity,
\beq
    \delta n_\rex(r)
     := \sum_{n_r,\ell}
        \lk n_{\rex,n_r\ell}(r) - n^{(0)}_{\rex,n_r\ell}(r) \rk.
\eeq
We find that the profiles of the order parameter and the quantum
depletion are similar.  This result reflects the fact that
the quantum depletion at a given position arises from zero-range
boson-boson repulsion and thus follows the local density of  the
condensate.
The impurity nearby induces a decrease (increase) in the
number of
atoms in the condensate by
$2.38\times10^{-3}$ ($2.40\times10^{-3}$)
and hence leads to a decrease (increase)
of the quantum depletion.  We remark that
in the far region from the impurity, i.e.,
beyond $r/d_\rB \sim 1.7$, the change in the condensate and depletion
due to the presence of
the impurity is tiny, but opposite in sign to that around the impurity.

Finally, we remark on the details of numerical calculations.
As pointed out in Ref.~\cite{Dalfovo2}, the convergence of the sum
in Eq.~(\ref{QD2}) is significantly slow.  We truncate the terms whose energy
exceeds $1000/\hbar\omega_{\rB}$.  Even with this truncation,
we can reproduce $90\%$ of the depletion obtained in the semiclassical
approximation \cite{Dalfovo2}, which implies that our approach is
quantitatively reasonable.

\subsection{Depletion vs.\ quantum numbers}
\begin{figure*}[ht]
\begin{center}
\includegraphics[width=15cm]{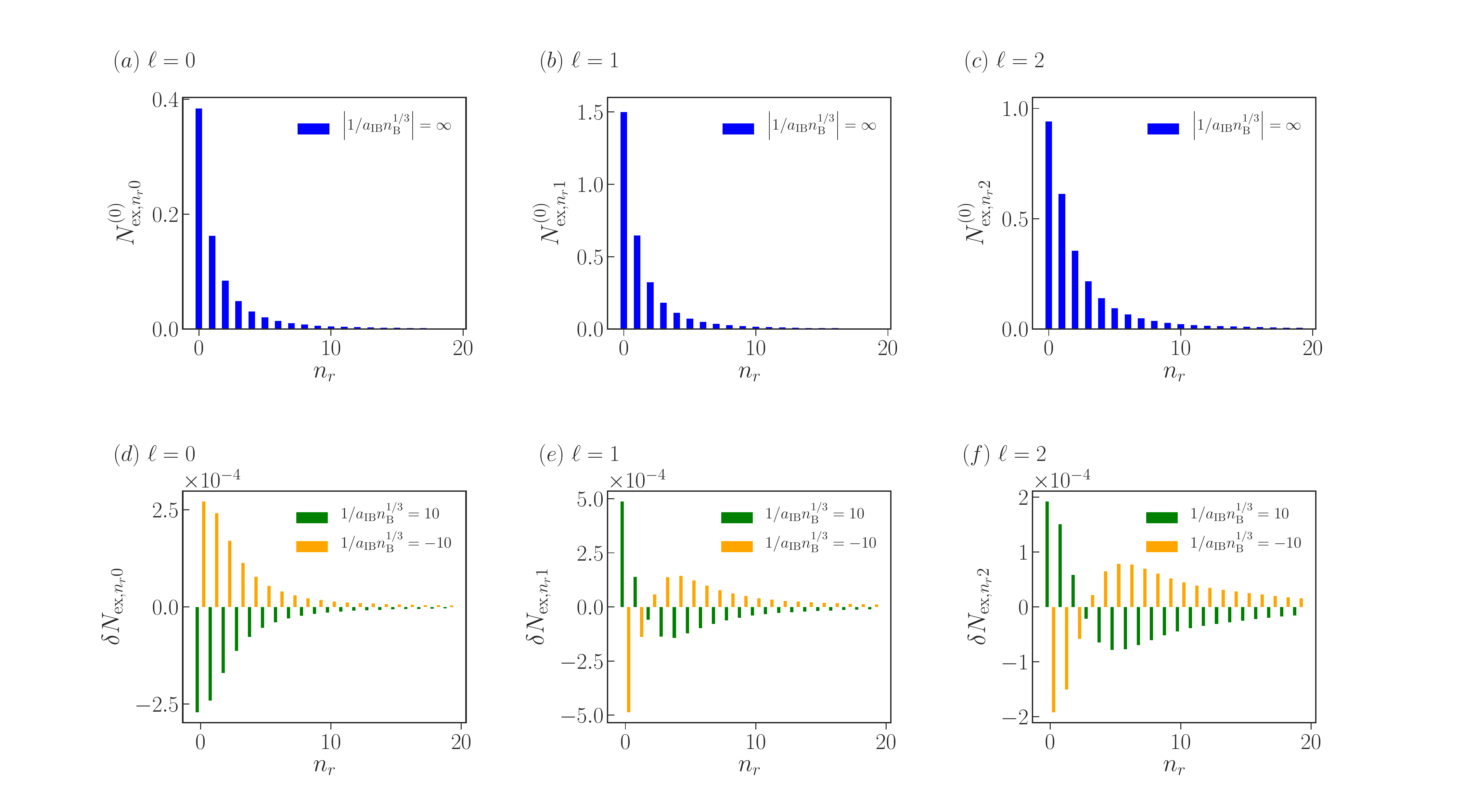}
\caption{
    (Color online)
    The number of the quantum depletion,
    $N_{\rex, n_r \ell}=\intr  n_{\rex, n_r \ell}(r)$, obtained
    as function of $n_r$ for
    $\ell=0,1,2$.  The impurity-free cases are depicted in (a)--(c),
    while the impurity-induced changes are in (d)--(f).
}
\label{fig:Nex_nl}
\end{center}
\end{figure*}

The $n_r$ dependence of the quantum depletion of the
condensate is shown for $\ell=0,1,2$ in Fig.~\ref{fig:Nex_nl}.
The figure indicates a qualitative difference
in the behaviors between $\ell=0$
and $\ell=1, 2$: When $g_\rIB>0$, the quantum depletion is
always suppressed by
$\ell = 0$ excitations with any $n_r$,
while it is suppressed (enhanced) by
high-lying (low-lying) excitations of $\ell \neq 0$.
When $g_{\rIB}<0$, the quantum depletion
behaves in the opposite direction to when $g_{\rIB}>0$.

\begin{figure*}[ht]
\begin{center}
\includegraphics[width=16cm]{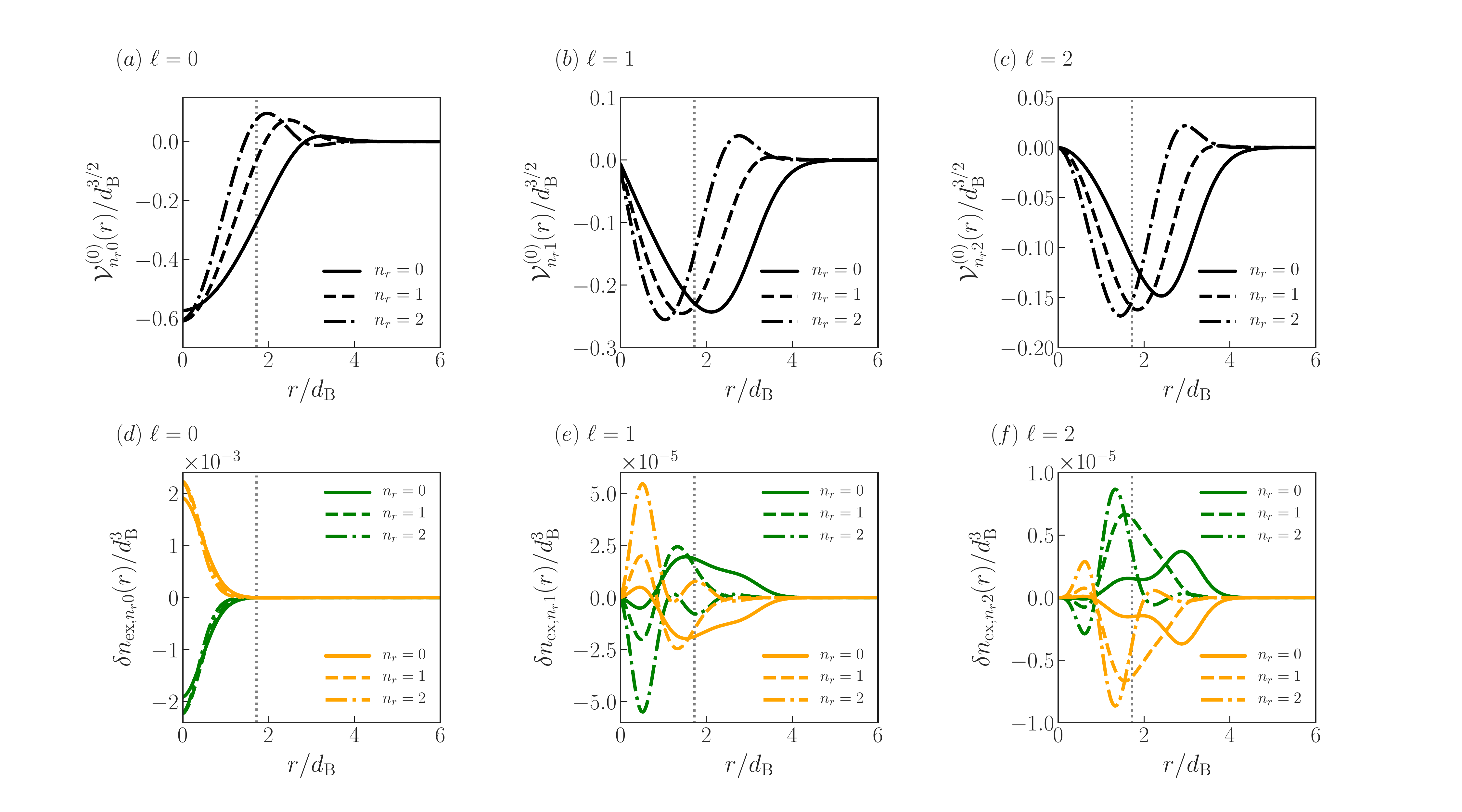}
\caption{
    (Color online)
    (a)--(c) Radial profile of the BdG eigenfunctions
     $\mathcal{V}_{n_r\ell}(r)$ in the absence of the impurity and
     (d)--(f) the corresponding change in the density of
     the quantum depletion due to the presence of the impurity.
    In (d)--(f), the green and the orange lines
    denote the $1/a_{\rm IB}n_\rB^{1/3}=+10$ and $-10$, respectively.
    The vertical dotted lines are drawn
    in the same way as in Fig.~\ref{fig:Veff}.
}
\label{fig:vnl}
\end{center}
\end{figure*}

To understand such behavior we illustrate the
BdG eigenfunctions $\mathcal{V}_{n_r \ell}(r)$
which correspond to ``hole'' excitations
in the absence of the impurity,
and differences in the quantum depletion at each state,
\beq
 \delta n_{\rex, n_r \ell}(r) :=
  n_{\rex, n_r \ell}(r) - n^{(0)}_{\rex, n_r \ell}(r),
\eeq
due to the presence of the impurity in Fig.~\ref{fig:vnl}.
We can see that for given $\ell$ and $n_r$, the corresponding change
in the density of depletion due to the presence of the impurity is
controlled by where the bottom of the dip of the BdG eigenfunction
$\mathcal{V}_{n_r\ell}(r)$ is located.

\begin{figure}[ht]
\begin{center}
\includegraphics[width=7cm]{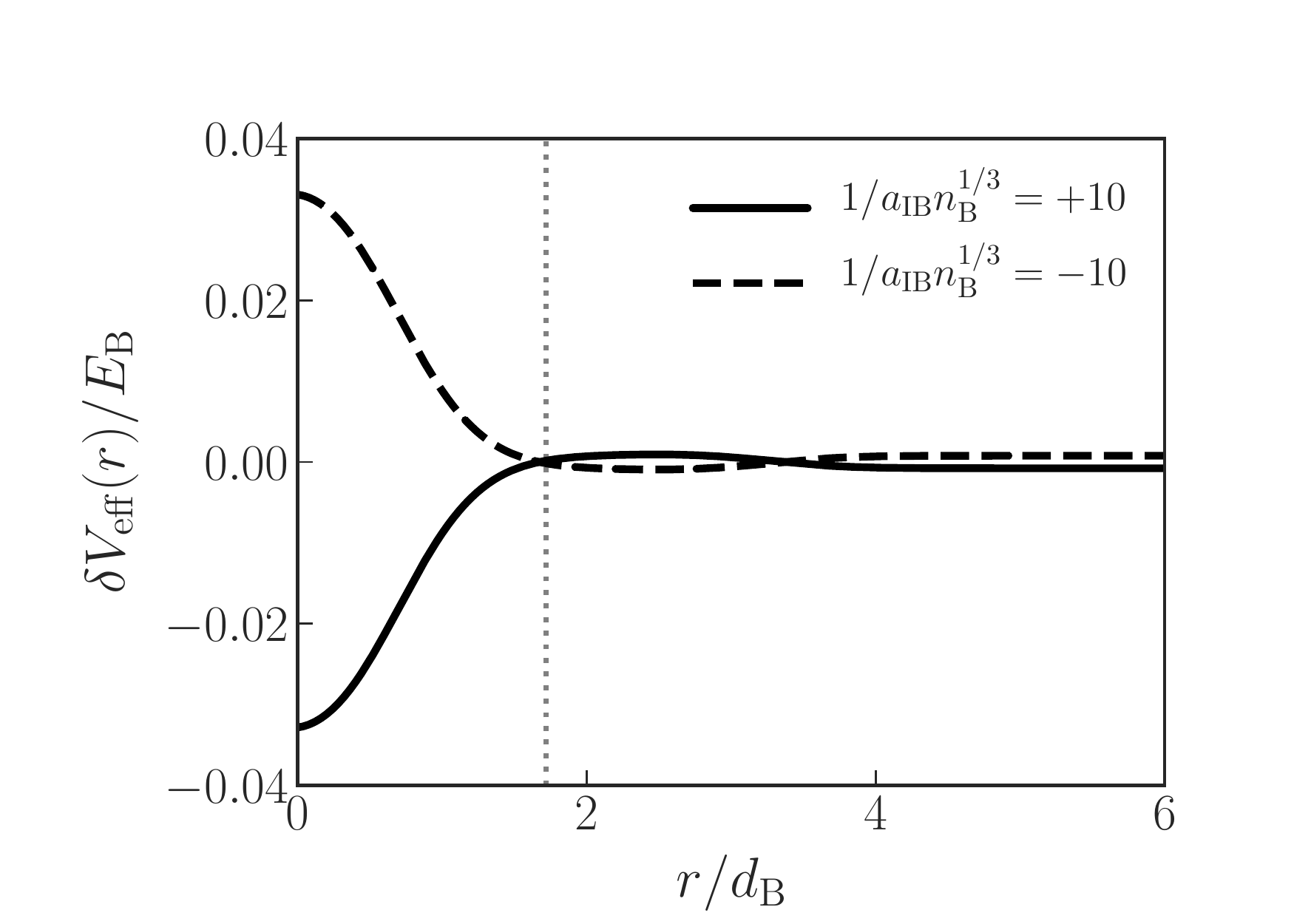}
\caption{
    (Color online) Difference of the radial profile of
    the effective potential nondimensionalized
    by $E_\rB = \frac{\hbar \omega_\rB}{2}$.
    The vertical dotted line
    denotes the point where
    the sign of $\delta V_{\rm eff}$ changes.
}
\label{fig:Veff}
\end{center}
\end{figure}

It should be noted that the behavior of the ``hole'' excitation
function $\mathcal{V}_{n_r, \ell}(r)$ is determined mainly
from the effective potential,
$
V_{\rm eff, \ell}(r)
    := V_{\rB}(r) + \frac{\hbar^2\ell(\ell+1)}{2r^2}
       + g_{\rm IB}\frac{\psi^2(r)}{4\pi}
       + g_{\rB}N_0 \frac{\Phi^2(r)}{2\pi} - \mu,
$
in the diagonal part of Eq.~(\ref{exc1}),
$
\mathcal{L}_{\ell}=-\frac{\hbar^2}{2m_{\rB}}
             \lk \frac{d^2}{dr^2} + \frac{2}{r}\frac{d}{dr} \rk
             + V_{\rm eff, \ell}(r).
$
We show the difference of the effective potential
in the presence of an impurity from that in its absence:
$\delta V_{\rm{eff}, \ell}(r):=V_{\rm{eff}, \ell}(r)-V^{(0)}_{\rm{eff}, \ell}$
in Fig.~\ref{fig:Veff}.

\begin{figure}[ht]
\begin{center}
\includegraphics[width=9cm]{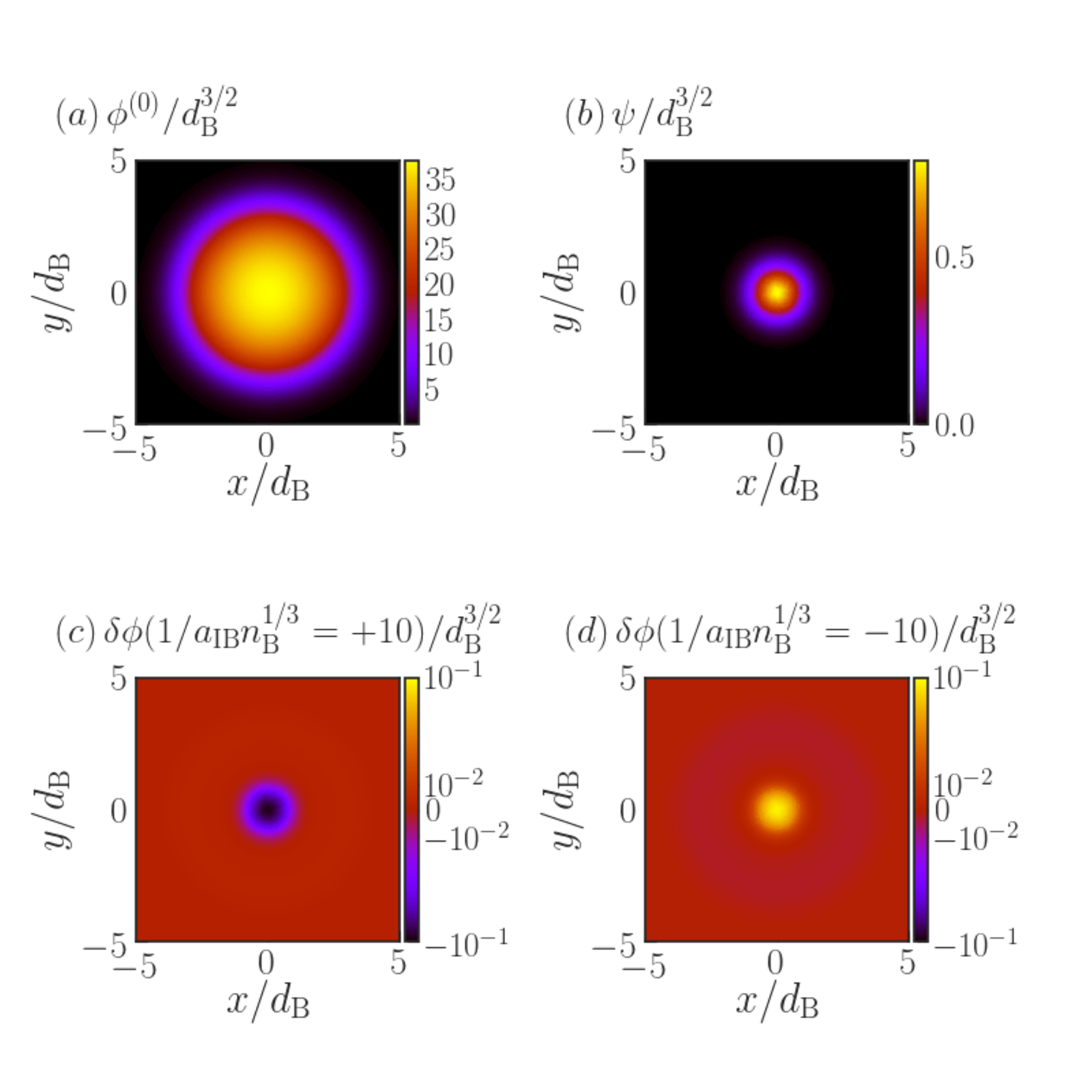}
\caption{
(Color online)
Real space profiles at $z=0$ of
(a) the order parameter $\phi_{\rB}(x,y,z=0)$ in the absence of
the impurity,
(b) the impurity's wave function $\psi_{\rI}(x,y,z=0)$, and
the differences of the order parameter
$\delta\phi_{\rB}(x,y,z=0)$
in the presence of the impurity with
(c) repulsive and (d) attractive boson-impurity interactions.}
\label{fig:phi_xy}
\end{center}
\end{figure}

\begin{figure}[ht]
\begin{center}
\includegraphics[width=9cm]{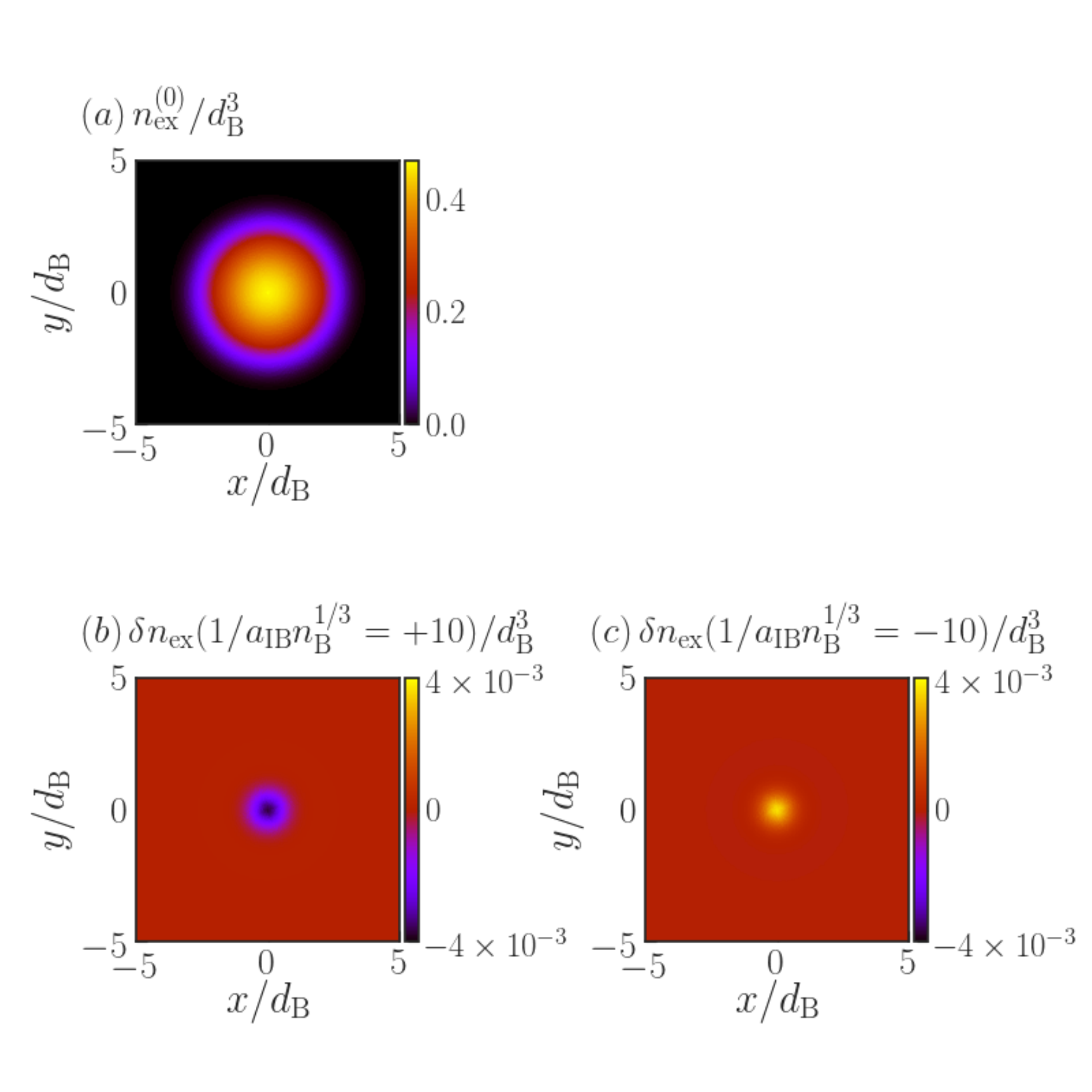}
\caption{
(Color online)
Real space profiles at $z=0$ of
(a) the density of depletion in the absence of
the impurity
$n_{\rm ex}(x,y,z=0)$ and
the differences of the density of depletion
$\delta n_{\rex}(x,y,z=0)$
in the presence of the impurity with
(b) repulsive and (c) attractive boson-impurity interactions.}
\label{fig:nex_xy}
\end{center}
\end{figure}

Figure~\ref{fig:Veff} shows that the $\delta V_{\rm{eff}, \ell}(r)$
directly reflects the difference of the radial profile of the condensate
(see Fig.~\ref{fig:order_parameter}(b)) and changes the sign
at $r/d_\rB\sim1.7$.
When the peak of $\mathcal{V}_{n_r, \ell}(r)$ comes near this point,
the sign of the quantum depletion changes.
In the case of $\ell=0$ in which no centrifugal potential occurs,
the dip bottom of $\mathcal{V}_{n_r\ell}(r)$
appears at the center of the effective potential,
as shown in Fig.~\ref{fig:vnl},
and the presence of the impurity gives a small isotropic change
to the potential around the center,
which thus leads to negative
or positive $\delta n_{\rex, n_r0}(r)$ therein.
On the other hand, when $\ell\neq0$,
the presence of the centrifugal potential allows the dip bottom of
$\mathcal{V}_{n_r\ell\neq0}(r)$ to move away from $r=0$ and,
for sufficiently small $n_r$
(corresponding to sufficiently high ``hole'' excitation level $-E_i$),
to enter the regime where the impurity-induced change
in the density of the condensate is opposite in sign,
while $\mathcal{V}_{n_r\ell\neq0}(0)=0$ is required for regularization.
Consequently,
the change in the depletion for $\ell\neq0$
appears inevitably away from the center
but within the influence of the impurity's wave function,
which can be summarized in terms of real space densities at $z=0$
in Figs.~\ref{fig:phi_xy} and \ref{fig:nex_xy}.
For these reasons, the depletion of the condensate
depends strongly on the excitation energy as well as the $\ell$ value.

We note that the depletion for $\ell=0$
is relatively small in comparison with that for $\ell=1,2$.
Since the integration (\ref{QD2}) includes
the Jacobian $\propto r^2$, the contributions from the amplitude
of $\mathcal{V}_{n_r, \ell=0}(r)$ dipped
around $r=0$ are suppressed, while those from the amplitudes of
$\mathcal{V}_{n_r, \ell\not=0}(r)$
around the dip bottoms remain unsuppressed.
This is why
the depletion for $\ell = 0$ is smaller than $\ell=1, 2$.
This situation is the same as the problems of quantum mechanics
in harmonic oscillators.

\section{Summary}
We have studied the quantum depletion of the condensate
in the trapped Bose-polaron
system and found that the net change of the depletion,
$\delta N_{\rex}
=\sum_{n_r, \ell}\intr\delta n_{\rex, n_r\ell}(r)$,
induced by the impurity can be negative (positive)
for the repulsive (attractive) impurity-boson interaction.
These qualitative results can be understood as follows:
1) We have solved a set of coupled equations for the condensate,
the impurity's wave function, and Bogoliubov excitations.
2) Among them, the first two are apparently coupled
with each other, but the condensate alone is affected by
the other because we have ignored the kickback effect on the
impurity by assuming that the impurity is tightly bounded in the trap.
3) Then, the resultant condensate, together with the impurity's wave function,
gives a small dip (bump) to the effective potential in the BdG
equation (see Fig.\ref{fig:Veff}),
which determines the spectra of excitations, i.e., the depletion.

The negative change of the depletion that we have found in the case
of the repulsive impurity-boson interaction in this study
seems counter intuitive, since in general the impurity-boson
interaction, no matter whether repulsive or attractive, disturbs the
condensation in uniform systems.
Thus, we consider this result as a specific
feature in trapped systems.
Moreover, this result does not depend on the shape of
trap potentials and can be observed, for example,
in a box trap potential.

Finally, we give a brief comment on possible experimental observations
of the results found in this study.  Although the
single-impurity-induced change in the number of depletion is minuscule, i.e.,
$|\delta N_{\rex}|\sim 10^{-3}$
for weak coupling considered here,
we expect that such change in the number of depletion is detectable
for a dilute impurity gas
even in weak coupling and more easily in strong coupling.
For instance, {\it in-situ} experiments
\cite{insitu7,insitu8,insitu9,insitu10,insitu11,insitu12}
could be used to obtain images of bosonic excitations,
while the photoabsorption spectroscopy by lasers \cite{Vor1,Vor2}
could select the azimuthal quantum number $\ell$ in the excitation
number $N_{\rex, n_r\ell}$.  Nevertheless, these experimental observations seem
challenging because of smallness of the depletion.
To enhance
the impurity-induced change in the quantum depletion, we propose
a method that uses bosonic impurities.
With this setting, the amplitude of the impurity wave function is increased;
therefore, the impurity-induced change in
the number of the quantum depletion can be enhanced.
For studies of such situations, we
will progressively extend the framework of our theory to deal with
the system with many impurities
, incorporating the many-body effects of impurities, e.g.,
the effective impurity-impurity interactions mediated
by medium fluctuations \cite{Cetina1, Nakano3}.

For another direction of research, we will further develop
the present method to look into
zero-mode physics \cite{Nakamura1, Nakamura2, Takahashi1},
few-body physics,
self-localization,
and so on.

\section{Acknowledgments}
We are grateful to the RIKEN symposium {\it Thermal field theory and its applications 2018} supported by iTHEMS, where the present work was initiated.
This work was supported in part by Grants-in-Aid for Scientific
Research through Grant Nos.~17K05445, 18K03501, 18H05406, and 18H01211,
provided by JSPS.


\begin{thebibliography}{99}
\bibitem{com2}

We do not discuss here the self-localization of impurities
\cite{Cucchietti1,Sacha1}.
The weak boson-impurity coupling constant used here
keeps the system from a possible Bose collapse
expected for strong attractive couplings in trapped systems \cite{Kalas1,Bruderer20}.
\end{thebibliography}


\end{document}